\documentclass[a4paper,11pt]{article}
\pdfoutput=1 

\usepackage{jinstpub} 
\usepackage{xcolor} 
\usepackage{subcaption}
\usepackage{siunitx}
\usepackage{fancyvrb}
\usepackage{soul} 
\usepackage{lineno} 
\usepackage{subcaption} 
\usepackage{float}

\title{The design and performance of CUBES -- a CubeSat X-ray detector}

\author[a,b,1]{Rakhee Kushwah,\note{Corresponding author.}}
\author[a,b]{Theodor A. Stana} 
\author[a,b]{Mark Pearce}

\affiliation[a]{KTH Royal Institute of Technology, Department of Physics, \\106 91 Stockholm, Sweden.}
\affiliation[b]{The Oskar Klein Centre for Cosmoparticle Physics, AlbaNova University Centre,\\106 91 Stockholm, Sweden.}

\emailAdd{rakhee@kth.se}

\abstract{
CUBES is a X-ray detector payload which will be installed on the KTH 3U CubeSat mission, MIST.
The detector comprises cerium-doped Gd$_3$Al$_2$Ga$_3$O$_{12}$ (GAGG) scintillators read out with silicon photomultipliers through a Citiroc Application-Specific Integrated Circuit. The detector operates in the energy range
$\sim$35--800 keV. The aim of the CUBES mission is to provide experience in the operation of these relatively new technologies in a high-inclination low earth orbit, thereby providing confidence for component selection in more complex satellite missions. The design of the CUBES detector is described, and results from performance characterisation tests carried out on a prototype of CUBES, called Proto-CUBES, are reported. Proto-CUBES was flown on a stratospheric balloon platform from Timmins, Canada, in August 2019. During the $\sim$12~hour long flight, the performance of Proto-CUBES was studied in the near-space environment. As well as measuring the X-ray counts spectra at different atmospheric depths, a 511~keV line from positron annihilation was observed.}


\begin{document}
\maketitle
\flushbottom

\section{Introduction}

\begin{figure}[ht]
  \centerline{\includegraphics[width=\textwidth]{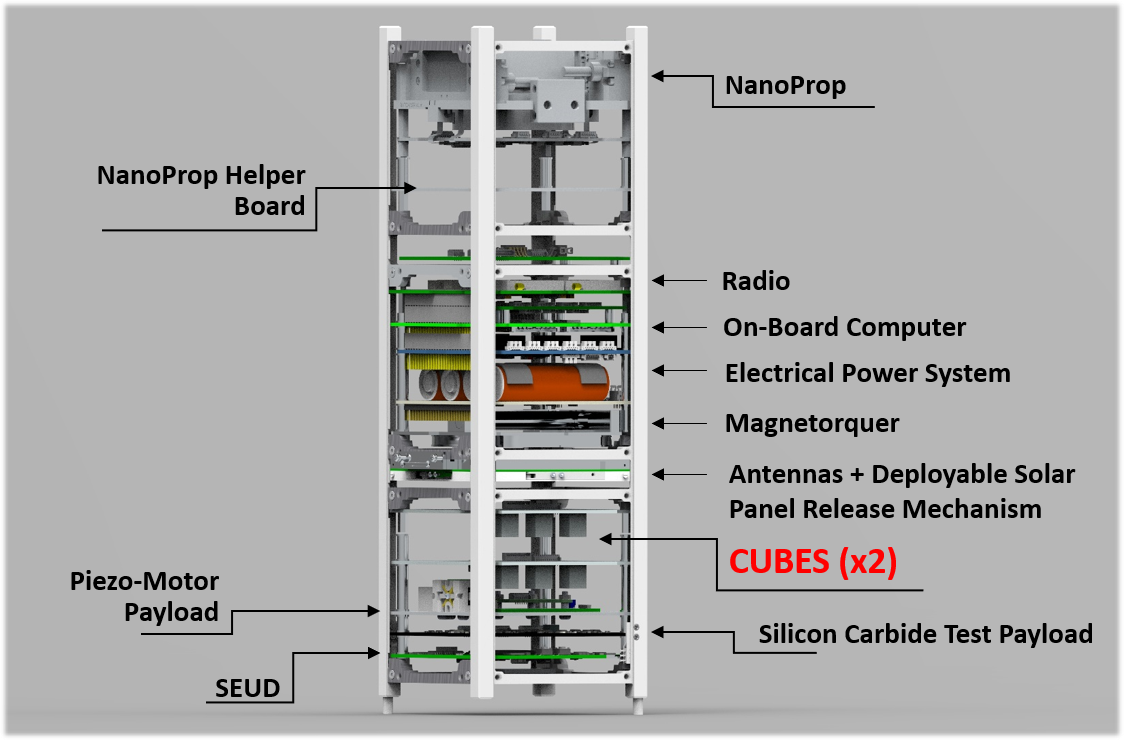}}
  \caption{A CAD image of the MIST satellite, which comprises three 10$\times$10$\times$10~cm$^3$ units. The top and bottom units contain the science payloads, while the central unit contains satellite sub-systems. The two deployable photovoltaic cell arrays, and four deployable UHF antenna are not shown. The sides of the satellite are partially covered in order to manage the thermal environment of satellite components. CUBES is mounted behind an aperture in these insulation materials.}
  \label{fig:mist-stackup}
\end{figure}

CubeSat missions have become a popular way to enhance science, technology, engineering and mathematics (STEM) subjects at universities.
Students gain first hand experience in the design, implementation and testing of electronic and mechanical space systems in a multidisciplinary and collaborative environment~\cite{cubesat}. The KTH Space Center~\cite{spacecenter} engages students in 
the construction of a 3U\footnote{The dimensions of each unit (U) are standardised to 10$\times$10$\times$10~cm$^3$} CubeSat mission, called MIST (MIniature Student saTellite)~\cite{MIST}, shown in Figure~\ref{fig:mist-stackup}.
The payloads are provided by research
groups at KTH and Swedish industry~\cite{MIST}, while satellite mechanics and
ancillary sub-systems (e.g. power, on-board computer, radio communications) are
procured from the company Innovative Systems in Space (ISISpace).
MIST is planned to be launched in 2022 into a sun-synchronous (98$^\circ$ inclination) orbit at an altitude of $\sim$640~km, with a local time of ascending node of 10:00.

The CUBES (CUBesat x-ray Explorer using Scintillators) payload is designed to measure the energy of incident X-rays in the 35--800~keV band. This is achieved using two $\sim$0.2U assemblies, each comprising cerium-doped Gd$_3$Al$_2$Ga$_3$O$_{12}$ (GAGG) scintillators read out with silicon photomultipliers through an Application-Specific Integrated Circuit (ASIC). 

The silicon photomultiplier (or multi-pixel photon counter, MPPC, as it is referred to in this paper) can replace traditional vacuum-tube photomultipliers (PMT) in scintillator detector systems. Although well-established in terrestrial applications~\cite{sipmt}, there is little experience of using the devices in the space environment.
The MPPC is a light-weight, compact, and robust solid-state single photon detector which is insensitive to magnetic fields. 
For a bias voltage of only $\sim$50~V, a gain similar to that obtained with a PMT operating at $\sim$1000~V can be achieved, for comparable quantum detection efficiency. 
A disadvantage is that the dark counting rate is relatively high and temperature-dependent. Moreover, it is found to increase with high-energy proton irradiation~\cite{mppc_irrad1,mppc_irrad2}. The temperature-dependence of the gain necessitates active bias voltage control.
GAGG scintillators are well suited for the detection of X-rays due to their high stopping power (density $\sim$6.6~g/cm$^3$) providing a mean free path of 0.34 mm and 0.98 mm for  X-ray photons with energy 40 keV and 100 keV respectively~\cite{nist1}. They are mechanically robust and non-hygroscopic, thereby simplifying packaging requirements. Their light-yield is high, $\sim$55~photons/keV, and comparable to CsI(Tl). The scintillation emission is centred at $\sim$520~nm, which is reasonably well-matched to the MPPC peak spectral sensitivity at $\sim$450~nm.  Studies~\cite{GAGGreview} have demonstrated a low internal background, that the light-yield dependence on temperature is small and not significantly affected by exposure to radiation, and that activation effects are small.   
The scintillation decay time is fast, $\sim$90~ns, and is preserved by the intrinsic fast response of the MPPC.
The resulting fast current pulse is processed by a low-power ASIC. 

One of goals of the CUBES mission is to assess the performance of these relatively new technologies in low-earth orbit in preparation for a possible future research mission, e.g. the gamma-ray burst polarimeter, SPHiNX~\cite{SPHINX}, which has completed a Phase A study in the Swedish {\it InnoSat} national satellite programme. 
Measurement data will also allow students to study the properties of the X-ray radiation present in low-earth orbit, as well as potentially allowing the observation of transient events connected to, e.g., solar flare events~\cite{solarflare} and gamma-ray bursts~\cite{GRB}. Data from CUBES are expected to be useful when interpreting results obtained by another MIST payload, SEUD (Single-Effect Upset Detector)~\cite{seud}, which aims to study mitigation methods for in-orbit single event upset rate in a field-programmable gate array (FPGA).

Other space missions have adopted GAGG scintillator coupled to MPPC read out, e.g. the GARI-1 and -2 missions expected to be launched to the International Space Station in 2021 and 2022, respectively, under the U.S. Department of Defence Space Test Program. Another example is GRID~\cite{GRID}, a Chinese 6U CubeSat mission launched in October 2018. The proposed HERMES mission~\cite{HERMES} uses a novel solution where GAGG scintillators are read out using a silicon drift detector.

University-built CubeSat missions are known to have a relatively low success rate~\cite{swartwout,aerospace6050054}. Due to cost and time constraints, missions are usually constructed using commercial off-the-shelf (COTS) components with little or no design redundancy. This is the case for CUBES. Comprehensive qualification testing is therefore an important aspect of ensuring mission success. As a part of the design qualification phase of CUBES, a prototype detector, Proto-CUBES, has been developed, primarily for functional and thermal testing. 
The qualification phase also includes an evaluation of measurements made during a stratospheric balloon flight, at altitudes of $\sim$20--30~km.  
Final qualification tests of the CUBES flight model will be conducted once the payload is integrated with the MIST satellite, and will include thermal-vacuum testing, shock/vibration tests, as well as functional performance tests. Total ionising dose studies of key components using a Co-60 source were planned, but the majority of tests have been indefinitely postponed due to the COVID-19 pandemic. The MPPC power supply module has been tested, however~\cite{MP_thesis}.  

In this paper, we present the design details of CUBES and the results obtained from characterisation studies and balloon flight of the Proto-CUBES instrument. The paper is organised as follows. The CUBES payload is outlined in Section~\ref{sec:cubes-instrument}. In Section~\ref{sec:calibration}, on-ground performance, calibration and thermal studies of Proto-CUBES are presented. Results from the Proto-CUBES balloon flight campaign are described in Section~\ref{sec:timmins-campaign}. An outlook on future developments is provided in Section~\ref{sec:outlook}.

\section{CUBES design overview}
\label{sec:cubes-instrument}

\begin{figure}[ht]
  \centerline{\includegraphics[width=\textwidth]{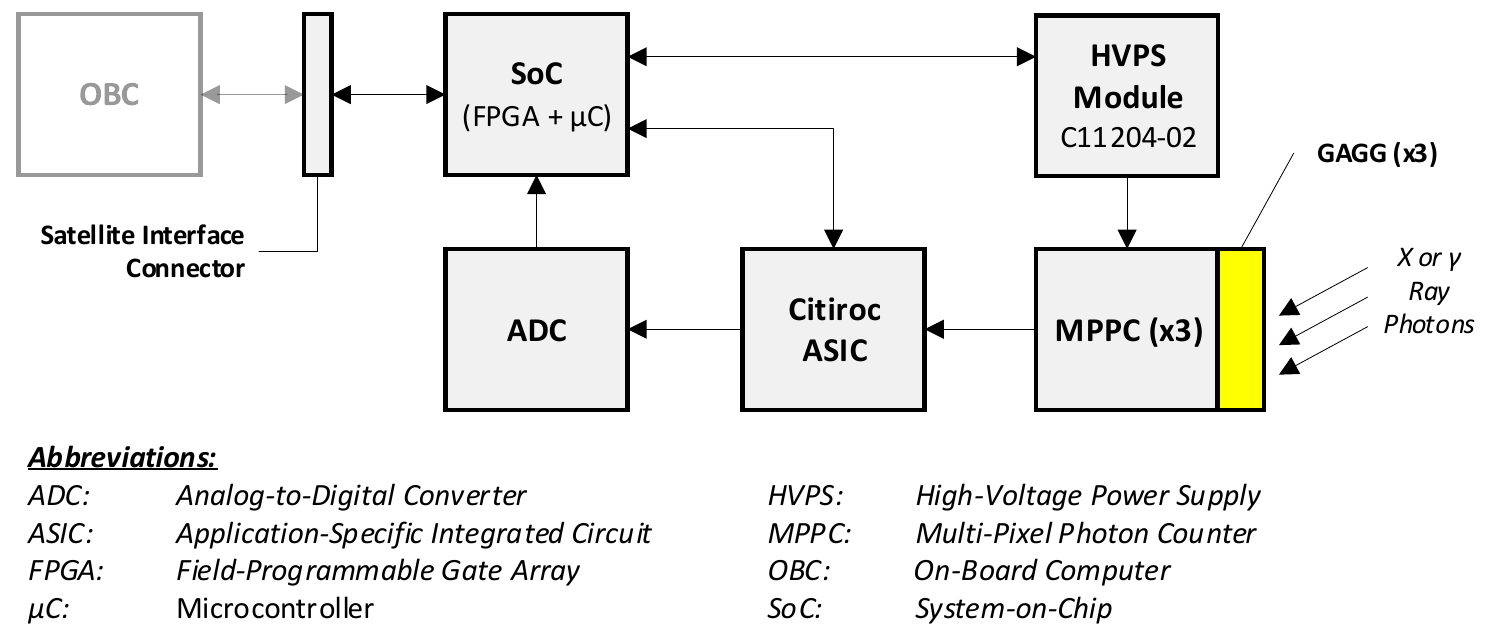}}
  \caption{CUBES block diagram.}
  \label{fig:cubes-block-diag}
\end{figure}

In the final configuration of the MIST satellite, there will be two CUBES
payloads (see Figure~\ref{fig:flightpcb}) in the bottom unit of the CubeSat, as shown in
Figure~\ref{fig:mist-stackup}. Each CUBES payload consists of a printed circuit
board (PCB) which contains all the instrument's elements. A block diagram of
the CUBES PCB is shown in Figure~\ref{fig:cubes-block-diag}. The GAGG
scintillators are procured from the C$\&$A Corporation. Each of the three GAGG 
scintillators has a volume of 1$\times$1$\times$1~cm$^3$ and is wrapped with
layers of PTFE tape to increase the scintillator light-yield. To
prevent ambient light from reaching the scintillators, they are encased in a
3D-printed mould filled with opaque silicone rubber potting compound (RTV627). Each GAGG scintillator is
optically glued to an MPPC of the type S13360-6050PE 
from Hamamatsu Photonics. A MPPC is an array of avalanche photodiodes (APDs) operated in
Geiger mode, which generates a short current pulse when a photon deposits a
sufficient energy in the scintillator material. Each APD is a sensitive
photodiode which multiplies the photocurrent under the action of a reverse bias
voltage. The bias voltage is provided by a programmable ``high-voltage'' power
supply (HVPS) module, the C11204-02, also from Hamamatsu Photonics. This device can
provide the very stable voltage needed by the APDs (on the order of tens of
$\mu$V) and it has built-in programmable  temperature compensation, which is
useful for compensating temperature-induced gain variations in the MPPCs.

MPPC current pulses are processed by the Citiroc ASIC manufactured by
Weeroc~\cite{citi-datasheet}. The Citiroc provides pre-amplification, shaping
and track-and-hold functions in separate high-gain (HG) and low-gain (LG)
chains, as well as fine adjustments to MPPC bias voltages via
digital-to-analog (DACs) on each of its 32 channels. Resulting pulses are
digitised on the CUBES PCB by means of a dual analog-to-digital converter
(ADC) chip, the 12-bit AD7356 from Analog Devices~\cite{ad7356-datasheet}.
The ASIC is controlled by logic embedded in a SmartFusion2 System-on-Chip (SoC) FPGA manufactured by Microsemi, part number
\mbox{M2S010-VFG256I}~\cite{fpga-datasheet}. The SoC also contains a Cortex-M3
microcontroller integrated on the same chip, which is used on CUBES to implement
communication to the MIST on-board computer (OBC) and control the C11204-02
HVPS module.

The amplitude of the current pulse output by the MPPC depends on
the deposited X-ray photon energy. To characterise the radiation environment
around the satellite, the data acquisition (DAQ) system implemented in the FPGA
of the CUBES SoC collects pulse-height spectra from each of the scintillator
detectors. The data are periodically stored in time-stamped histograms. 

MPPC pulses with an amplitude exceeding the ASIC trigger threshold, above which current pulses arriving
from the MPPCs are converted to voltages and sent to the ASIC's output. The ADC
converts these into a digital representation which is used by the CUBES DAQ
system to decode the memory address to read and increment a counter
value. Six histograms are stored by the CUBES DAQ system, each of the three
MPPCs is allocated one histogram for the HG chain and one for the LG chain.
Each of the six histograms contains 2048 bins (one bin contains two ADC values).

Following each DAQ run, a DAQ data file is stored on CUBES which contains the
six histograms and a 256-byte header. The header contains ``housekeeping'' data
such as DAQ run time, instrument live time, temperature, MPPC bias voltage and
current drawn by the MPPC -- a key indicator of radiation damage to the MPPC,
since the dark current has been shown to increase with prolonged radiation
exposure~\cite{mppc_irrad1} -- and number of pulses from the
MPPCs crossing the configurable threshold on the ASIC. Header included, CUBES
stores 24832 bytes of data to its memory during one DAQ run.

\section{Characterisation and calibration studies}
\label{sec:calibration}

\begin{figure}[ht]
    \centerline{\includegraphics[width=.75\textwidth]{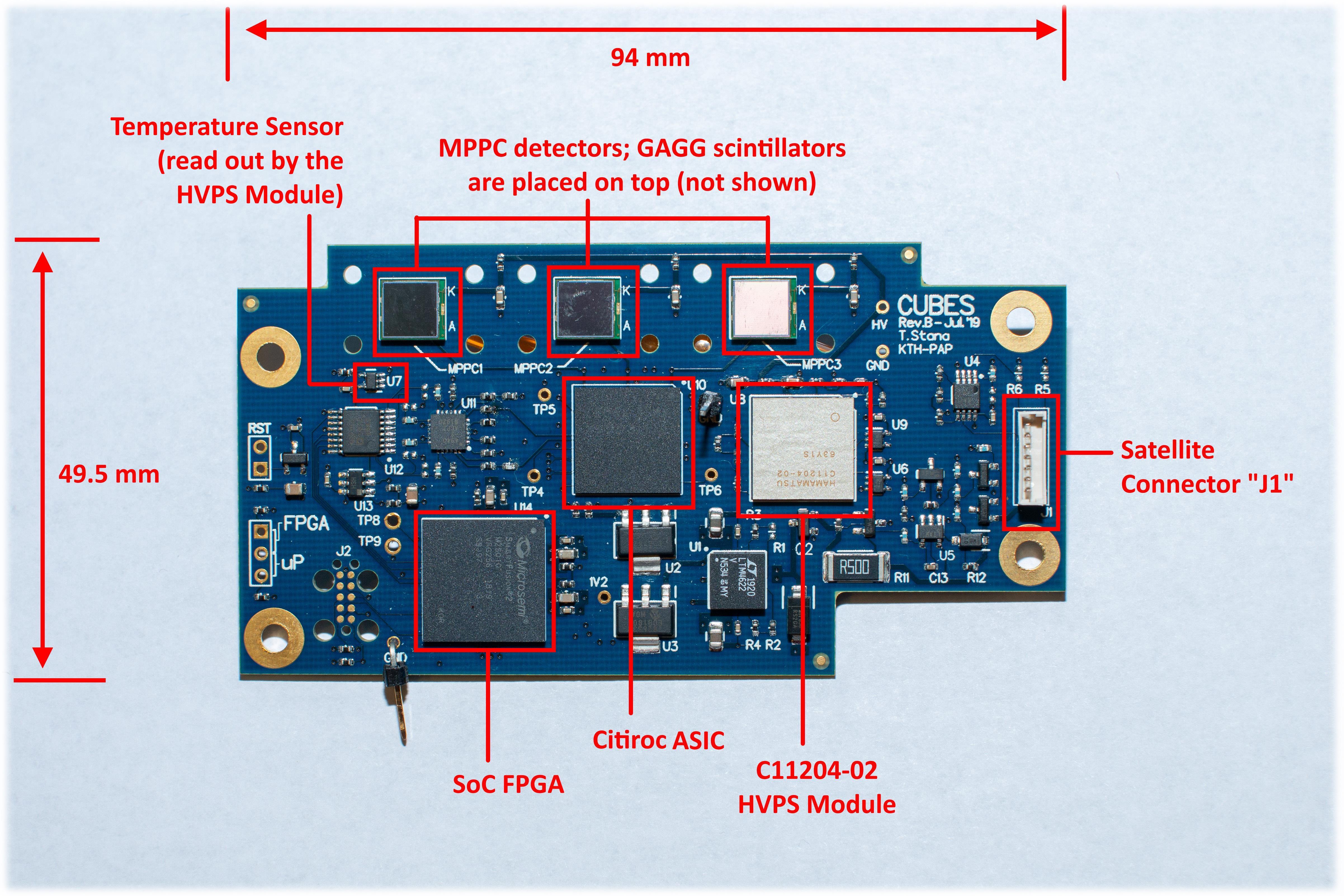}}
    \caption{\label{fig:cubes-pcb-revB} The Proto-CUBES printed circuit board.}
\end{figure}

\begin{figure}[ht]
\centerline{\includegraphics[width=\textwidth]{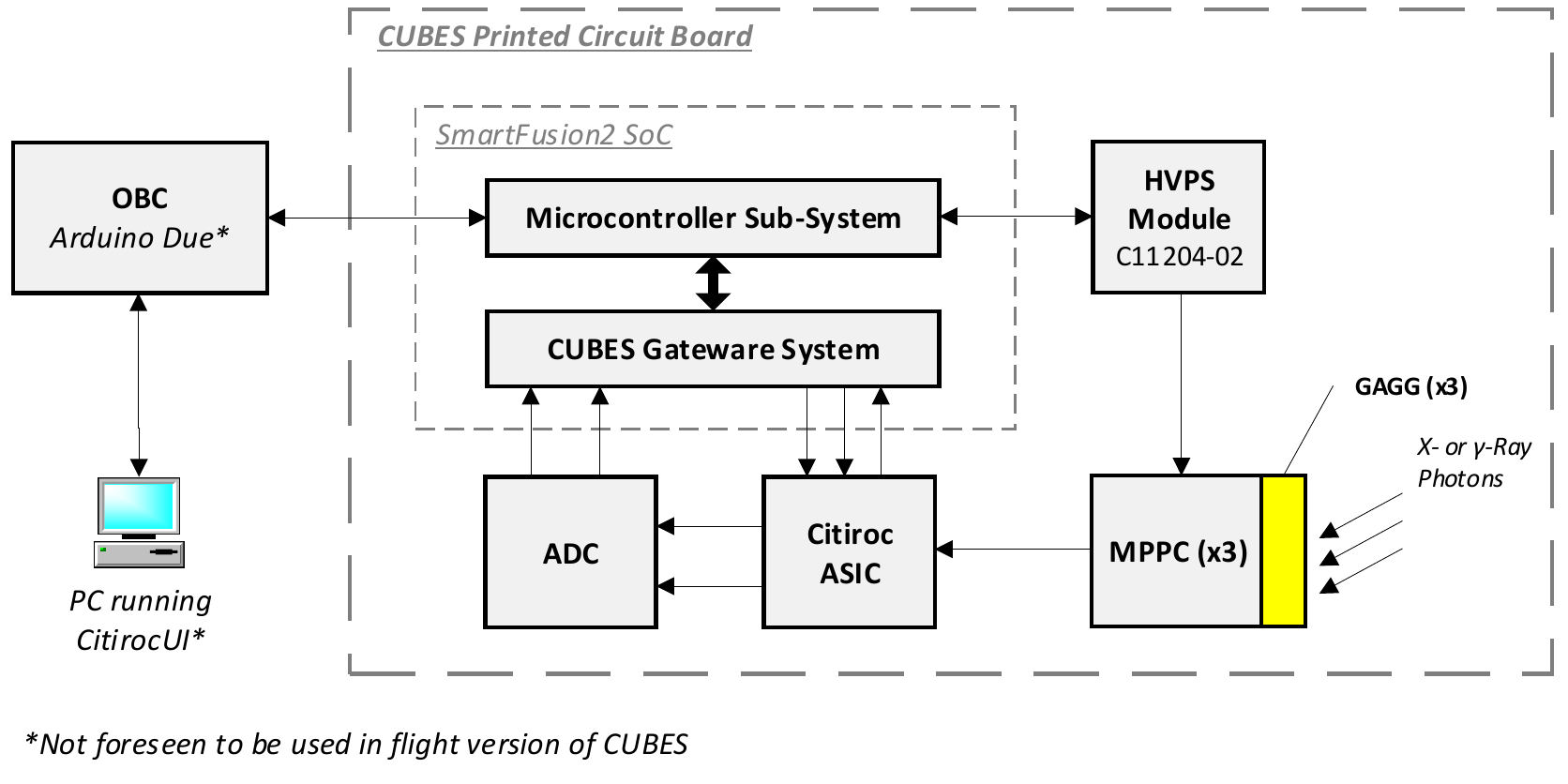}}
\caption{\label{fig:cubes-block-diag-dev-setup} The Proto-CUBES system setup.}
\end{figure}

The Proto-CUBES PCB implementation of the system depicted in Figure~\ref{fig:cubes-block-diag} is shown in Figure~\ref{fig:cubes-pcb-revB}. 
During development of the experiment
and characterisation studies, an Arduino Due is used as an OBC simulator, as
depicted in Figure~\ref{fig:cubes-block-diag-dev-setup}. The
simulator implements the MIST Space Protocol (MSP) as the data link
layer protocol in the seven-layer OSI model~\cite{osi-model-ieee}; the physical
layer protocol is the Inter-Integrated Circuit (I$^2$C) bus. On CUBES,
MSP commands are used to configure the ASIC and the DAQ system, and retrieve
scientific and housekeeping data.

To retrieve the data from CUBES for analysis, a modified version of the CitirocUI software is used. 
The CitirocUI software is provided by the ASIC vendor for use with their ASIC evaluation boards. 
For this project, the CitirocUI software has been modified by us to send commands over a serial
port to the Arduino. The Arduino translates most of these commands into MSP
commands to configure the ASIC and the CUBES DAQ system. During  in-orbit
operations of CUBES, this modified CitirocUI will be used to test ASIC configurations before uploading them to CUBES.
After ASIC and DAQ configurations are applied, most of the in-orbit operations
of CUBES are foreseen to be autonomous, with data  files downloaded from the payload periodically.

For Proto-CUBES, each GAGG scintillator has a volume of 5$\times$5$\times$5~mm$^{3}$ and is
wrapped in several layers of 0.075~mm thick PTFE tape to improve the collection
efficiency for scintillation photons. Each GAGG scintillator is coupled to a MPPC
using optically transparent Epotek~301 glue. The ensemble of GAGG scintillators and MPPCs is covered
with an opaque mouldable black silicone rubber material (generically referred to
as "Sugru") to ensure no stray light enters the detector ensemble.

\subsection{Rate characterisation using an LED setup}
\label{sec:rate}
\begin{figure}[ht]
    \centerline{\includegraphics[width=.75\textwidth]{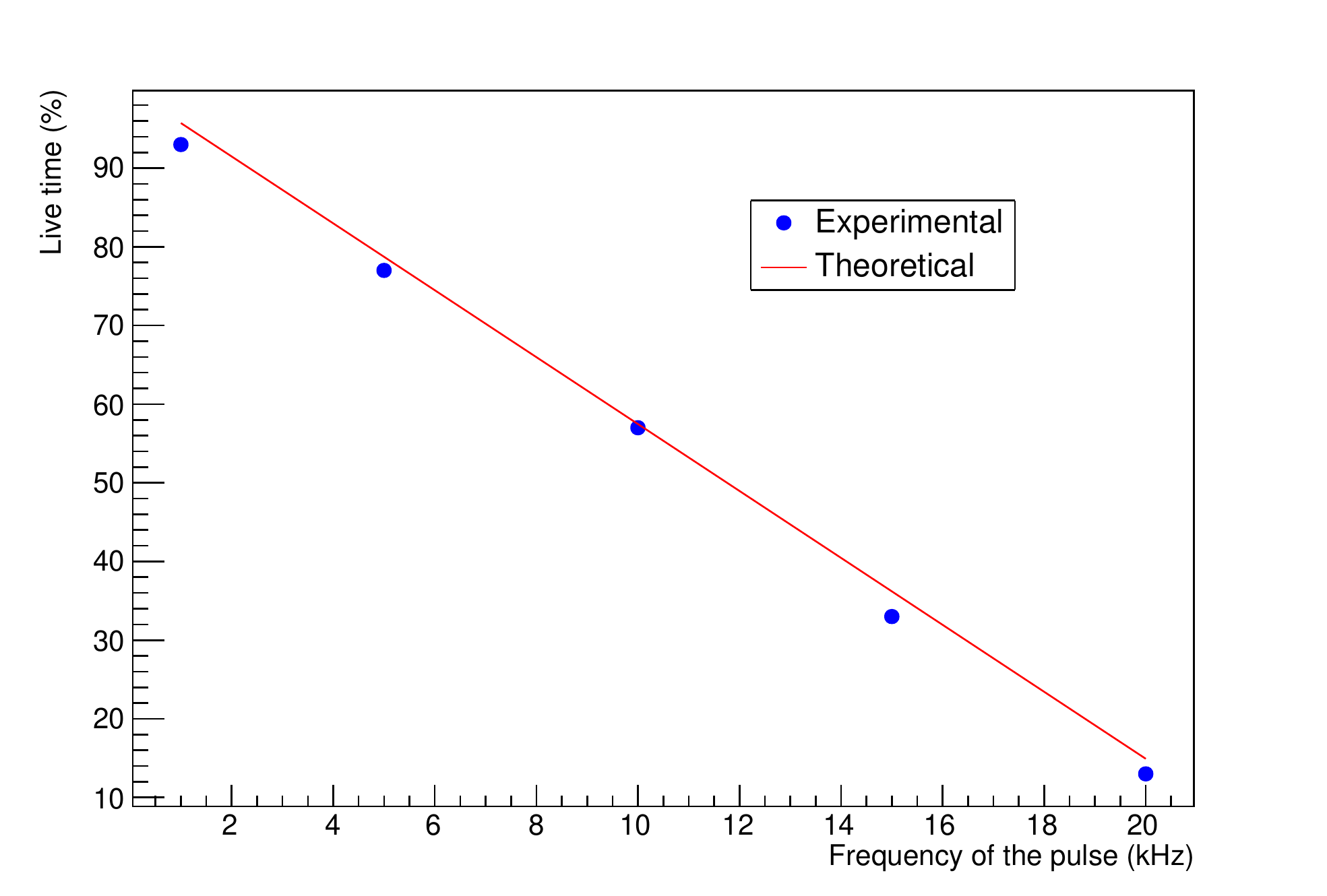}}
    \caption{\label{fig:rate} The effect of LED pulsing frequency on the measured live time.}
\end{figure}

The CUBES trigger rate will strongly depend on orbital position.
High rates are expected in the polar regions and at the South Atlantic Anomaly (SAA). The live time is dictated by the time taken by each pulse to be processed by the Citiroc ASIC, which for CUBES is \mbox{$\sim$43~$\mu$s}. During this time, the CUBES DAQ system is unable to process subsequent pulses.
The live-time dependence on trigger rate has been studied using 
a fibre-coupled blue LED, driven by a pulse generator. The LED light is coupled to each of the MPPCs using an optical fibre splitter. Each fibre was terminated with a ST connector which mated with a receptacle mounted on the MPPC.
The LED pulse parameters were chosen to produce a MPPC current pulse which simulated a 59.5~keV energy deposit (Am-241 emission) in a GAGG scintillator. 
For each LED pulsing frequency, the live time was obtained from the housekeeping data in the DAQ data file. 
Figure~\ref{fig:rate} shows that the experimental live time values agree well with the theoretical expectation, defined as

\begin{equation}
    \mathrm{live\, time\, (\%)} = \frac{R_{i} \times T_{ASIC}}{T_{DAQ}}\times 100
\end{equation}
where $R_{i}$ is the number of incoming pulses, $T_{ASIC}$ = 43~$\mu$s as stated above and $T_{DAQ}$ is the total acquisition time. 

Using the energy spectra of trapped protons and electrons from SPENVIS~\footnote{\url{https://www.spenvis.oma.be/}} and a simplified GEANT4~\cite{geant4} model of CUBES, the maximum trigger rate in-orbit has been estimated. During passages through the South Atlantic Anomaly and in the polar region, the trigger rate is of the order of a few MHz, and measurements will be dead-time limited. During approximately 65\% of the orbit, the trigger rate is $<$10 kHz, however. 

\subsection{Energy characterisation using radioactive sources}
\label{sec:energy-range}

\begin{figure}[ht]
    \centerline{\includegraphics[width=.75\textwidth]{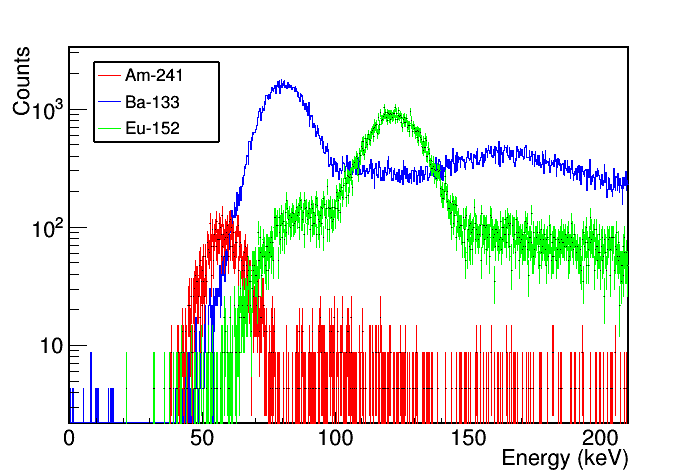}}
    \caption{\label{fig:HG}Energy spectra for radioactive sources acquired with the HG amplifier.}
\end{figure}

\begin{figure}[ht]
    \centerline{\includegraphics[width=.75\textwidth]{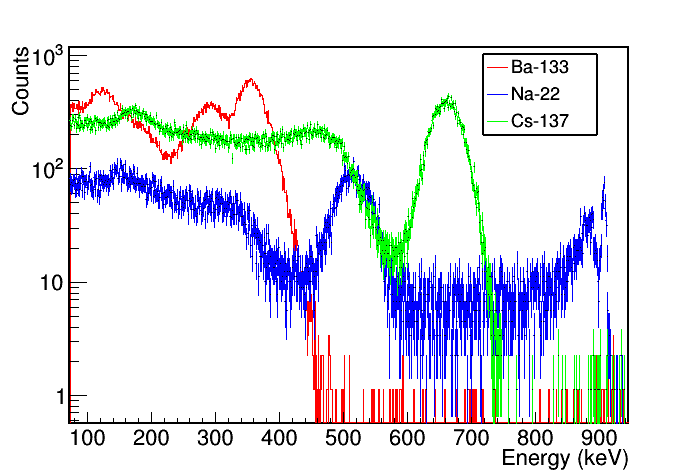}}
    \caption{\label{fig:LG} Energy spectra for radioactive sources acquired with the LG amplifier.}
\end{figure}

\begin{figure}[ht]
    \centerline{\includegraphics[width=.75\textwidth]{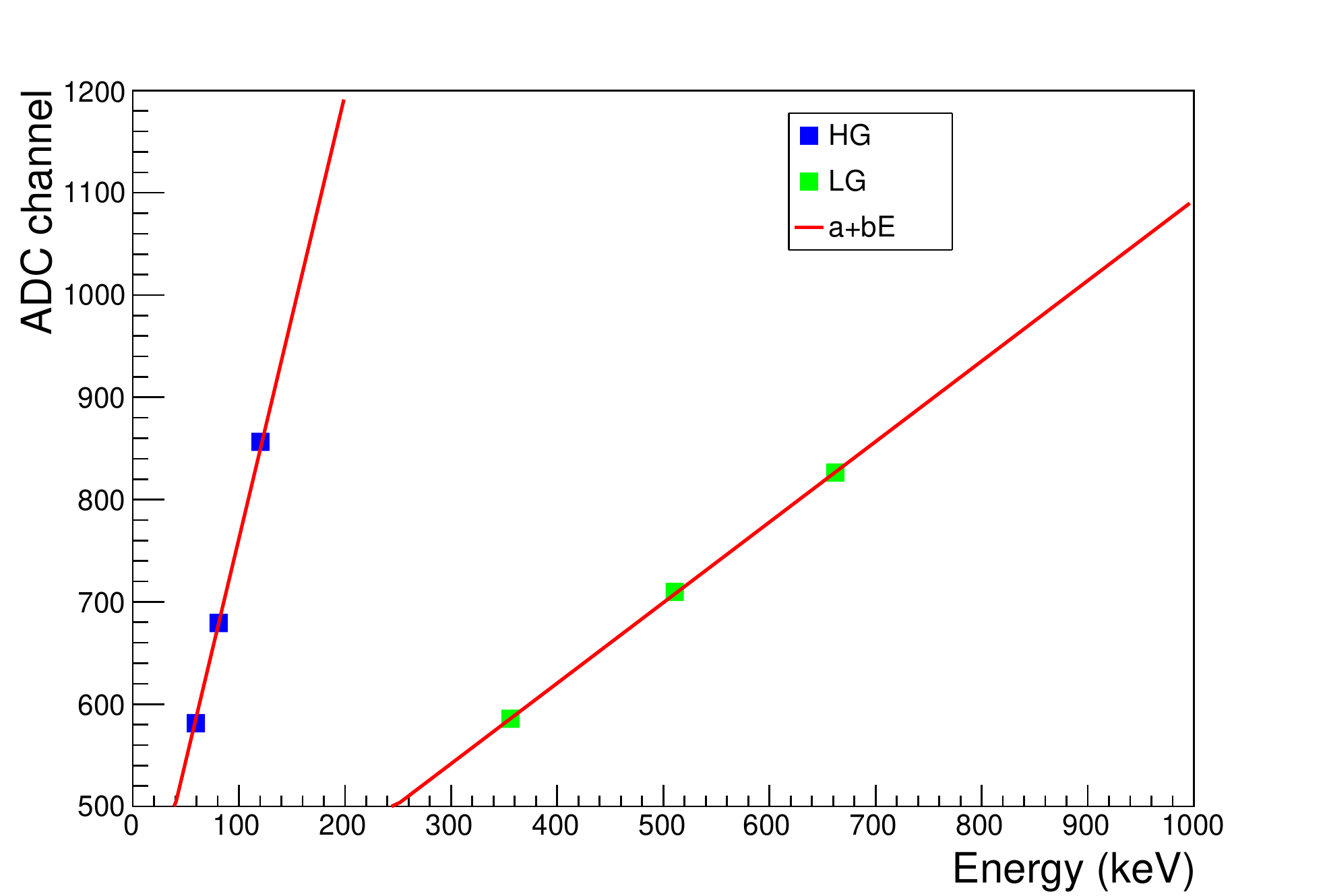}}
    \caption{\label{fig:linearity}Energy linearity of the HG and LG amplifiers, evaluated by irradiating a single GAGG scintillator with different radioactive sources. The error bars are smaller than the data point markers. The fit parameters, a (in ADC channels) and b (in ADC channels/keV) for HG chain are 327$\pm$1 and 4.342$\pm$0.010 respectively. For LG chain, a and b are 306$\pm$0.437 and 0.787$\pm$0.001 respectively.}
\end{figure}

\begin{figure}[ht]
    \centerline{\includegraphics[width=.75\textwidth]{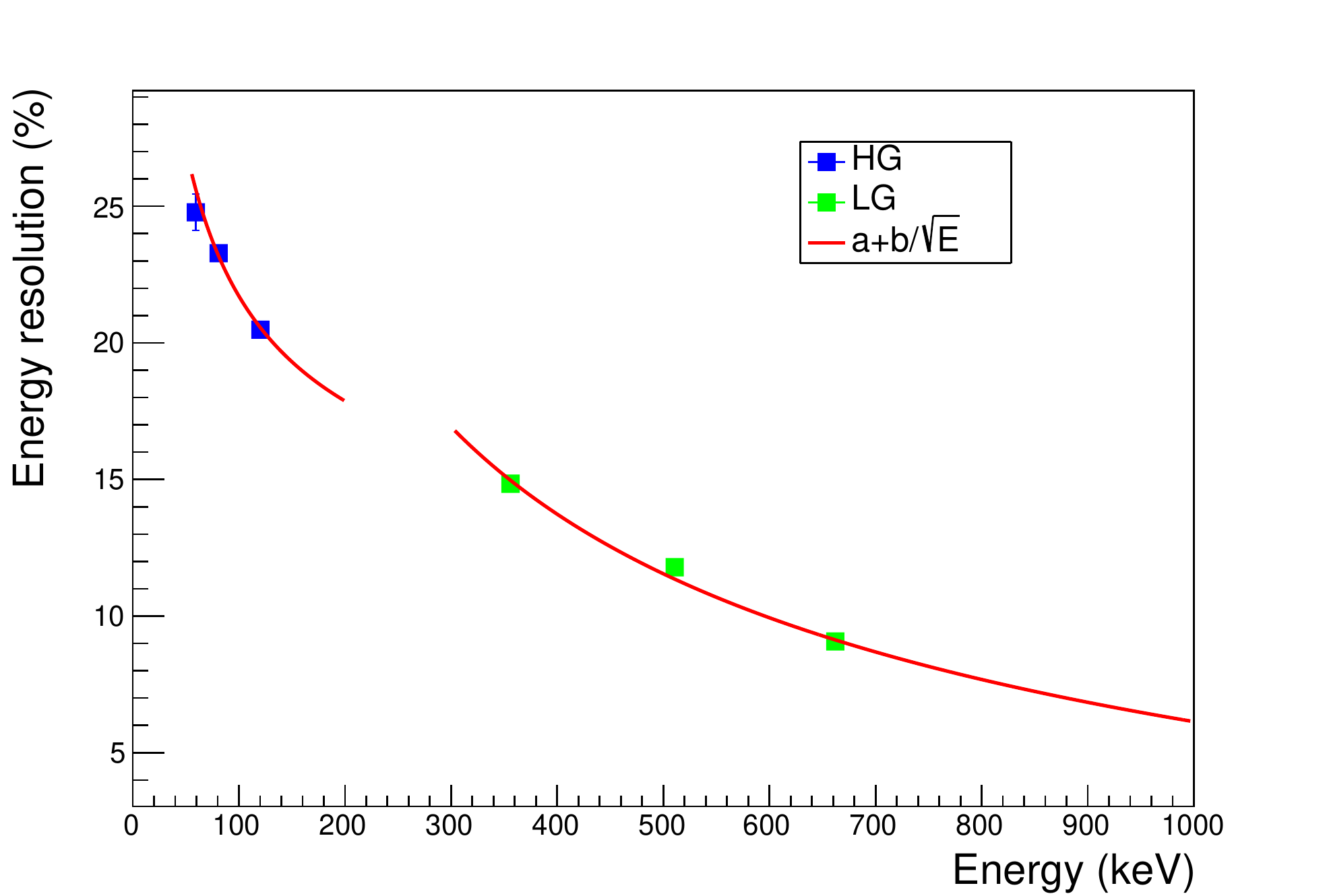}}
    \caption{\label{fig:resoln} The dependence of energy resolution on deposited energy for a representative GAGG scintillator. The majority of the error bars are smaller than the data point markers. The fit parameters, a and b for HG chain are 0.085$\pm$0.011 and 1.312$\pm$0.104. For LG chain, a and b are -0.069$\pm$0.006 and 4.131$\pm$0.145.}
\end{figure}

The detector setup was characterised to obtain the energy-channel calibration using different radioactive sources. The spectra obtained by the HG and the LG amplifier chains of the Citiroc for one of MPPC channels is shown in \mbox{Figure~\ref {fig:HG}} and \mbox{Figure~\ref {fig:LG}} respectively. The HG chain allows to measure 59.5 keV photons from Am-241, 81 keV photons from Ba-133 and 122 keV photons from Eu-152. The spectra from 356 keV photons from Ba-133, 511 keV from Na-22 and 662 keV photons from Cs-137 were acquired using the LG chain. The photopeak position shows a linear dependence on energy, as shown in Figure~\ref{fig:linearity}. The ASIC does not display significant non-linearity, as reported previously~\cite{MPPC-Citiroc-Pol}. All three GAGG scintillators showed a similar light output and energy-channel relation. Using the same spectra, the energy resolution is also computed and shown in Figure~\ref{fig:resoln}. The points are fitted with a function of the form, $f(E)= (a+b/\sqrt{E})$, as expected from the Poisson counting statistics.

With the lower trigger threshold set above the noise floor, the energy range achieved for each MPPC channel (referred to as MPPC1, MPPC2 and MPPC3 in the text) is shown in Table~\ref{tab:energy}. An upper threshold is also applied to avoid pulses which saturate the ADC connected to output of the Citiroc. There is a discontinuity in the energy ranges obtained from HG and LG chains. Moreover, differences in MPPC gain leads to different maximum energy values. For the flight detector, it is foreseen that these effects can be mitigated by adjusting the gain of HG and LG chains and/or the MPPC gain. 

\begin{table}[ht]
\centering
\caption{\label{tab:energy} Achieved energy range for each MPPC channel.}
\smallskip
\begin{tabular}{|c|c|c|c|}
\hline
Amplifier chain&MPPC1&MPPC2&MPPC3\\
\hline
HG &32 -- 156 keV &40 -- 200 keV &30 -- 140 keV\\
\hline
LG & 176 -- 856 keV  &225 -- 1174 keV &138 -- 763 keV\\
\hline
\end{tabular}
\end{table}

\subsection{Thermal characterisation}
\label{sec:thermal}
The temperature of CUBES is primarily dictated by sun illumination, and therefore depends on orbital position.
The MIST satellite is attitude controlled, and will be oriented such that CUBES payload will not be directly illuminated by the sun. 
The thermal design of MIST aims to provide an operating temperature range for CUBES of -20 -- 30~$^{\circ}$C. This temperature range is driven by the MPPC specifications, as well as the expected bias voltage tuning range achievable using the Citiroc's input DACs (see Section~\ref{sec:cubes-instrument}). 
Since the MPPC gain and the GAGG scintillator light-yield are temperature dependent~\cite{GAGGreview,gaggtemp}, temperature characterisation tests of the detector system have been performed. 

\begin{figure}[H]
    \centerline{\includegraphics[width=.75\textwidth]{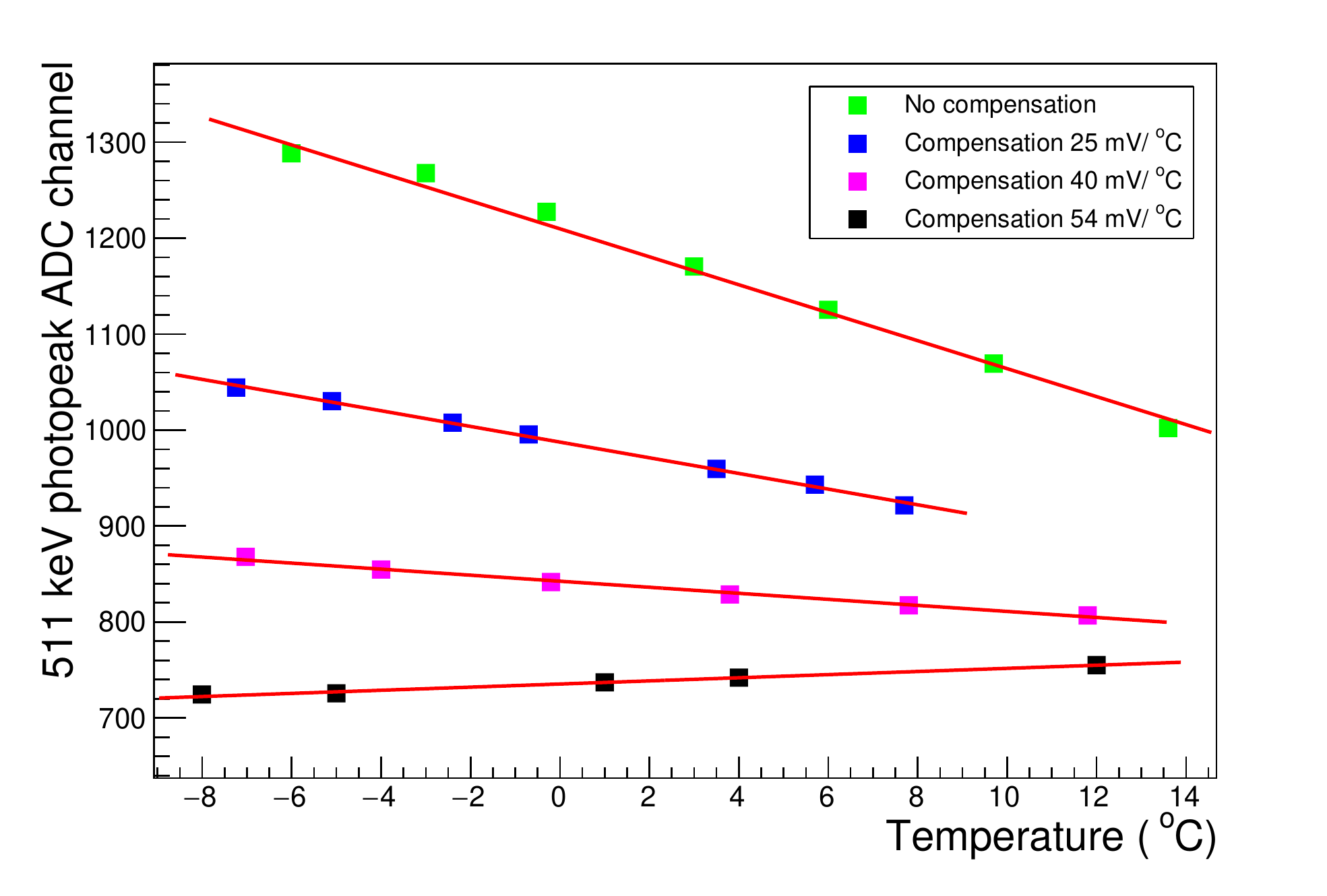}}
    \caption{\label{fig:thermal} The variation in the 511 keV photopeak position with temperature, T, for different temperature coefficients. The data comes from the MPPC2 channel, and is fitted with a linear function, $a+bT$ (red line). The fitted parameters are listed in Table~\ref{tab:fit_par_fig10}. The error bars are smaller than the data point markers. The recommended temperature coefficient for the MPPC, 54 mV/$^{\circ}$C, does not fully remove the temperature dependence.}
\end{figure}

\begin{figure}[H]
    \centerline{\includegraphics[width=.75\textwidth]{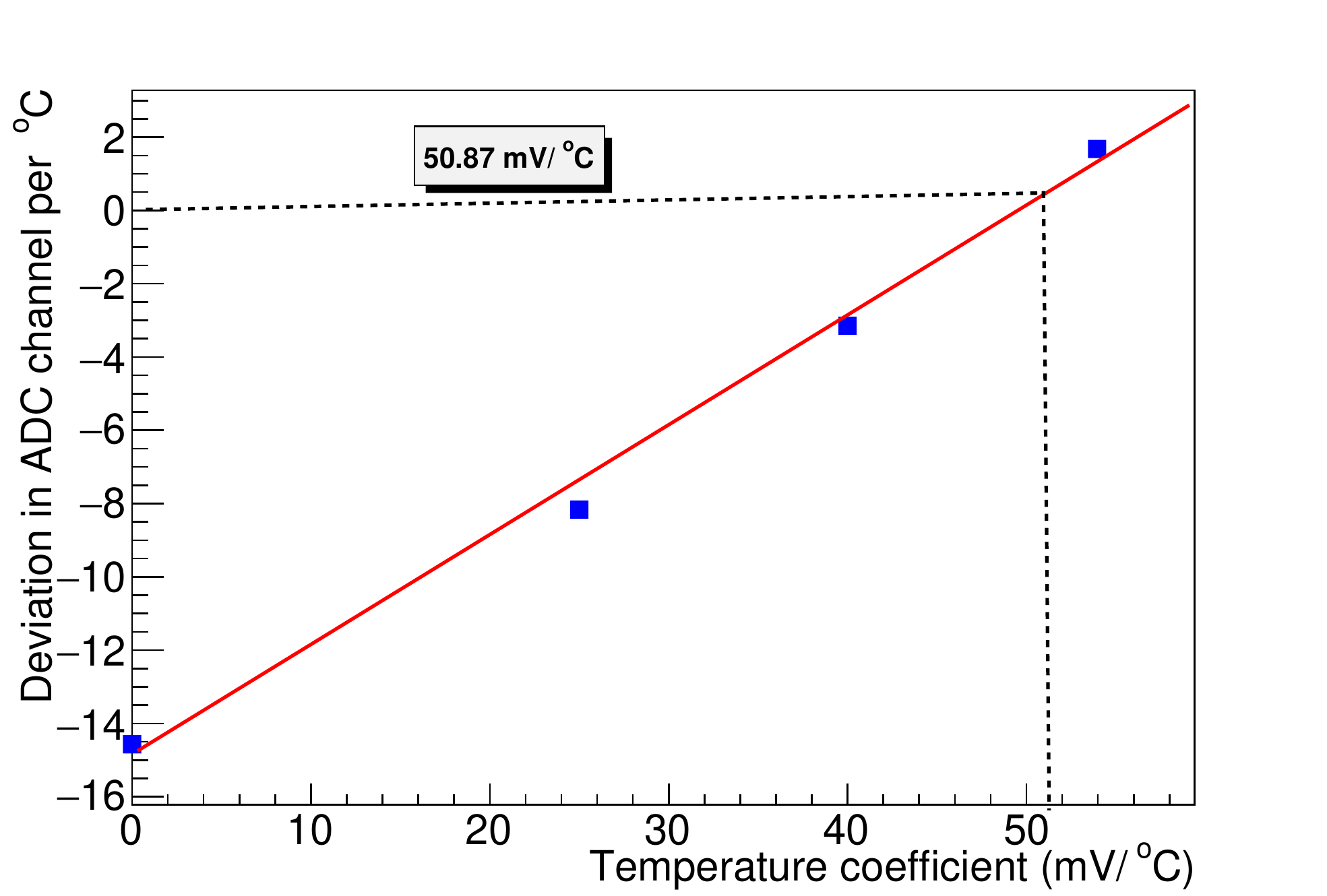}}
    \caption{\label{fig:slope} The temperature dependence of the 511~keV photopeak position
    (ADC channels/$^{\circ}$C) plotted against the applied device temperature coefficient, $T_{c}$. The error bars are smaller than the data point markers. The red line is a linear fit, $a+bT_{c}$, to the data points. The fitted parameters, a and b are -14.729$\pm$0.023 and 0.289$\pm$0.001.}
\end{figure}

As described in Section~\ref{sec:cubes-instrument}, the HVPS module, C11204-02, has a built-in temperature compensation circuit. The MPPC bias voltage is varied according to a device-specific temperature coefficient, thereby keeping its gain constant. The temperature coefficient of the MPPCs used here is 54 mV/$^{\circ}$C~\cite{hamamatsu}. To test this, the detector was placed inside a temperature controlled chamber, and the GAGG scintillators were irradiated with 511~keV photons from a Na-22 source. 
Spectra were acquired across the temperature range -10$^{\circ}$C to +15$^{\circ}$C, with each DAQ run lasting for 4~minutes. The variation in the position of the photopeak with temperature is shown in Figure~\ref{fig:thermal}. With the recommended temperature compensation of \mbox{54 mV/$^{\circ}$C}, the photopeak position exhibited a small temperature dependence.
Other studies have observed similar behaviour~\cite{gaggtemp}. The test was repeated using a range of temperature coefficients, and the results are shown in Figure~\ref{fig:thermal}. 
The temperature dependence of the peak position is characterised with a linear function. The dependence of the function slope (ADC channel/$^{\circ}$C) on the device temperature coefficient factor (mV/$^{\circ}$C) is shown in Figure~\ref{fig:slope}. The temperature dependence of the 511~keV photopeak position can be removed using a temperature coefficient of $\sim$50.9 mV/$^{\circ}$C.

\begin{table}[H]
    \centering
    \caption{\label{tab:fit_par_fig10} Fit parameters of $a+bT$ in Figure~\ref{fig:thermal}.}
    \smallskip
    \begin{tabular}{|c|c|c|}
    \hline
    Temperature Coeff. &$a$&$b$\\
    ((mV/$^{\circ}C)$)&(in ADC channel)&(ADC channels/$^{\circ}$C)\\
    \hline
    No compensation&1210$\pm$0.18&-14.57$\pm$0.02\\
    \hline
    25&987$\pm$0.18&-8.17$\pm$0.04\\
    \hline
    40&843$\pm$0.21&-3.14$\pm$0.03\\
    \hline
    54&735$\pm$0.16&1.63$\pm$0.02\\
    \hline
    \end{tabular}
\end{table}

\section{Performance tests on a stratospheric balloon flight}
\label{sec:timmins-campaign}
%
\subsection{Balloon flight environment}

Stratospheric balloon flights, operating at altitudes between 30--40 km, provide a means to acquire operational experience of instruments in the near-space environment.
Payloads may be subject to a similar range of temperatures to that found in orbit, and the mbar-level vacuum means that the primary heat transfer mechanisms are radiation and conduction. 
The balloon floats over the majority of the earth's atmosphere. Fluxes of cosmic particles are only partially attenuated providing a more realistic multi-component radiation environment than on-ground. 
Instruments are, furthermore, operated remotely over radio links, as is the case in orbit.
The flight duration can be between several hours to several weeks depending on the balloon type, launch season, launch site and flight trajectory. Payloads, in the absence of an unforeseen landing in, e.g., water, are usually
recoverable soon after the flight is terminated allowing inspection of hardware, and design changes prior to possible retesting.

The design of Proto-CUBES is relatively simple. The scintillator field-of-view is large, with no collimation, and there is no anticoincidence shield. The instrument therefore responds to energy deposits from the range of cosmic radiation present in the atmosphere. 
The nature of the radiation experienced by balloon-borne instruments depends on altitude. Galactic ('primary') cosmic-rays ($\sim$90\% p, $\sim$10\% He, $<$1\% other nuclei and electrons) interact with atoms at the top of the atmosphere ($\sim$40~km), creating showers of secondary particles which follow the direction of the momentum vector of the primary particle. While traversing the atmosphere, the charged and neutral secondary particles decay and lose energy, resulting in an altitude-dependent intensity of particles, with the maximum intensity known as the Pfotzer-Regener maximum~\cite{regener}.  

Part of the flux of secondary particles comprises X-rays photons which are generated through electromagnetic and nuclear processes in the atmosphere. Sources include $\pi^0$ decay, e$^{\pm}$ bremsstrahlung, e$^+$e$^-$ annihilation, and the de-excitation of nuclei arising from spallation/neutron-capture. Photon energies are degraded through multiple Compton scattering interactions, and the lowest energy photons are depleted through photoelectric absorption on atoms in the atmosphere.
The intensity of the upward-/downward-going components of the X-ray flux varies with altitude. The albedo flux is dominant.
In the Proto-CUBES energy range, the differential flux (cm$^{-2}$s$^{-1}$keV$^{-1}$sr$^{-1}$) depends on energy, E, as $\sim$E$^{-2}$~\cite{Dean,Petersen}. A flux of primary X-ray radiation (the cosmic X-ray background, CXB) impinges the top of the atmosphere.  
The nature of the flux is well-measured~\cite{cxb1}. Above 20~keV, the energy dependence of the differential flux is $\sim$E$^{-3}$ until $\sim$1~MeV. 
The spectrum at balloon altitudes is subject to atmospheric attenuation~\cite{CXB_atten}. Even at the highest altitudes, the CXB contribution is small compared to atmospheric X-rays.

The balloon flight took place in Ontario, Canada, corresponding to a geomagnetic latitude ($\lambda$) of $\sim$57.65$^\circ$, and a vertical cut-off rigidity of $\sim$1~GV~\cite{cutoff}.
Bremsstrahlung from precipitating electrons can increase the X-ray flux in the atmosphere during periods of elevated geomagnetic activity, especially in high-latitude locations. This contribution primarily generates ionisation effects at altitudes of $\sim$60~km~\cite{Bazilevskaya}. This source is not expected to affect our measurements, since no significant geomagnetic activity was reported during the flight~\cite{geomagnetic}.
The contribution from primary cosmic-ray electrons is suppressed due to the relatively high cut-off rigidity at the launch site. Secondary electrons are produced in the atmosphere. However, simulations studies conducted for different float altitudes ($h$) and geomagnetic latitudes show that the electron flux is negligible compared to the photon flux ($h$=30~km, $\lambda$=11.5--17.2$^\circ$)~\cite{Sarkar}, ($h$=38~km, $\lambda$=65.4$^\circ$)~\cite{XL-Calibur} .

Neutrons are copiously produced in the atmosphere as a result of weak interaction processes or nuclear processes such as cascades or evaporation. The resulting broad range of neutron energies, $\sim$0.1~MeV--10~GeV is further extended to thermal neutron energies as neutrons scatter off atmospheric nuclei~\cite{pogolino}. 
There may also be a local component created as cosmic rays interact with the gondola and payload structures. Depending on the detector material used, neutrons can be a significant source of background for X-ray instruments operated in the atmosphere.

\subsection{Instrument design and gondola integration}

\begin{figure}
\begin{center}
    \includegraphics[width=0.80\linewidth]{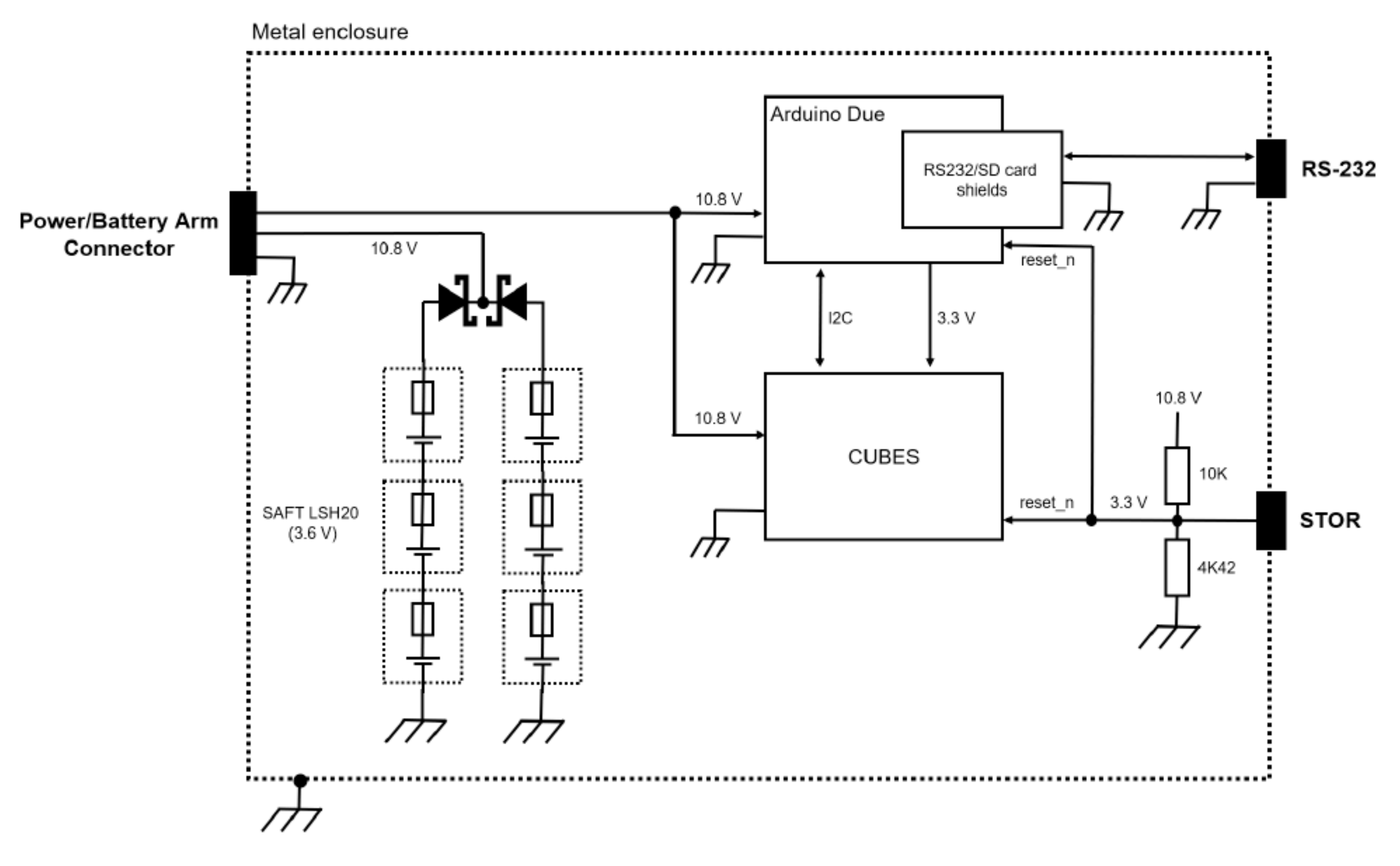}
\end{center}
\caption{A schematic diagram showing how the Proto-CUBES PCB (Figure~\ref{fig:cubes-pcb-revB}) is integrated into a self-contained system for the stratospheric balloon flight. The Proto-CUBES detector is controlled by an Arduino Due microcontroller, which is connected to the ground-station via a virtual RS232 link implemented over radio. An independent reset signal, "STOR", is also provided. The in-flight power system comprises high-power lithium-thionyl chloride batteries. During testing on the ground, the instrument can also be powered from a bench-top power supply using the "power/battery arm" connector. For flight, the internal battery circuit is activated by inserting an arming connector.}
\label{fig:balloon_payload} 
\end{figure}

A schematic diagram showing how Proto-CUBES is integrated into a self-contained system for the stratospheric balloon flight is shown in Figure~\ref{fig:balloon_payload}. 
Following the approach used for on-ground testing, the MIST OBC is simulated by an Arduino Due microcontroller. 
There is a command and data link to the Arduino via a virtual RS232 connection established over a radio link. 
Data is also stored on a SD memory card, which can be recovered post-flight.
An independent connection (``STOR'') is provided as a radio-controlled ``open-collector'' signal (active low), which allows a reset signal to be issued to the instrument during flight. The reset signal was used in cases where the instrument entered an unknown state, to revert the Arduino and CUBES microcontrollers, as well as the CUBES FPGA logic, to a known, ``start-up'' state. The instrument was powered by three 3.6~V SAFT LSH20 lithium-thionyl chloride batteries wired in series. Each battery has an in-built 5~A fuse. Two sets of serially connected batteries are diode-coupled in parallel to provide redundancy. The instrument consumes 150~mA, resulting in a power consumption of 1.65~W. One set of three batteries provides an operational time of at least 24~hours at -20$^\circ$C. Once armed the power cannot be remotely cycled, but a reset signal can be issued if a component becomes unresponsive.

\begin{figure}
\begin{center}
    \includegraphics[width=\linewidth]{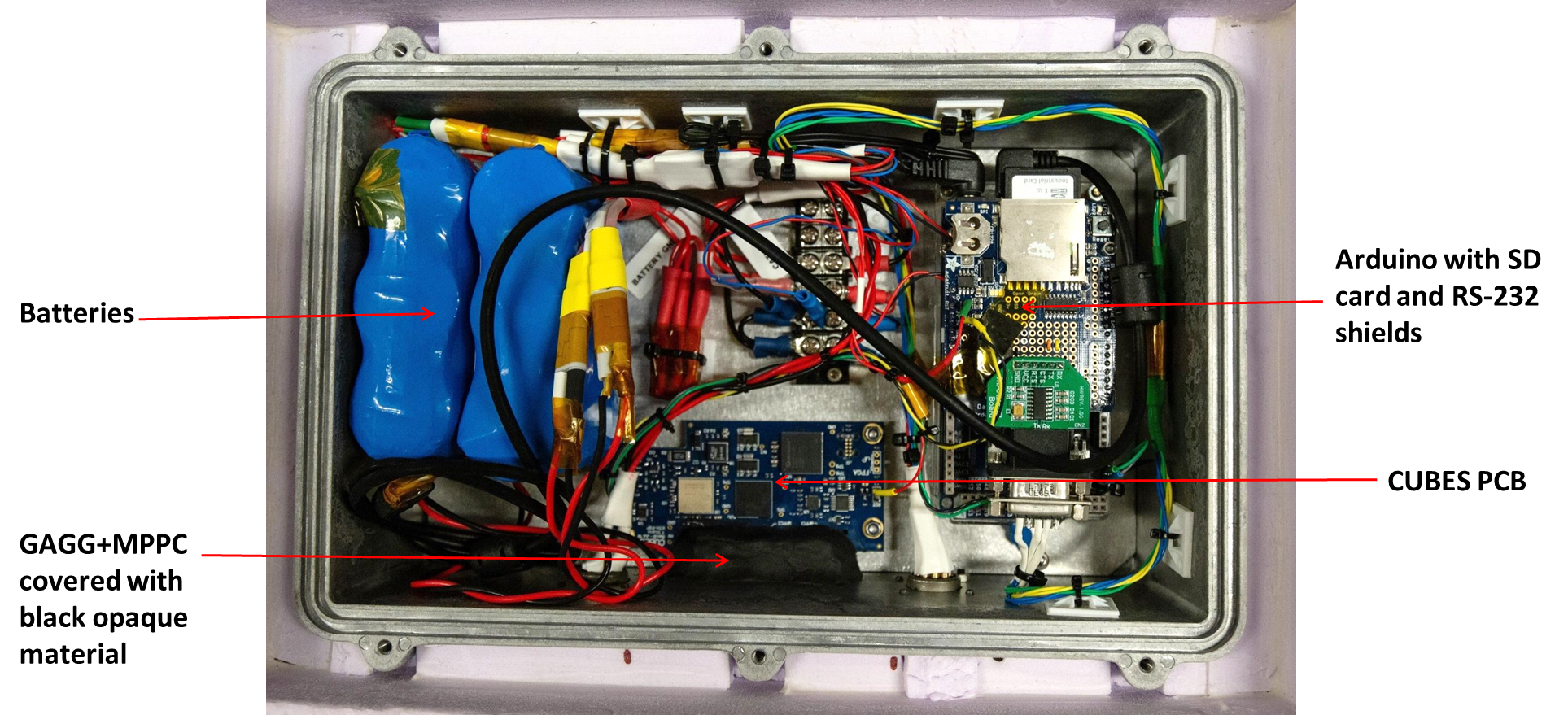}
\end{center}
\caption{The instrument enclosure (30$\times$20~cm, 10~cm tall) encased in styrofoam insulation.}
\label{fig:box} 
\end{figure}

The instrument components were placed in an aluminium enclosure with 2.5 mm thick walls, as shown in Figure~\ref{fig:box}. 
The side, top and bottom faces of the enclosure in the vicinity of the GAGG scintillators was reduced to a thickness of 1~mm.  
The transmission factor for a beam of X-rays with energy 20, 50, 200, 500~keV traversing 2.5~mm (1~mm)  thick aluminium is 
10\% (40\%), 78\% (91\%), 92\% (97\%), 95\% (98\%), respectively~\cite{NIST}. The enclosure was covered in 3~cm thick styrofoam sheets to provide thermal insulation.
The resulting assembly was wrapped in 200~$\mu$m thick aluminiumised mylar to reflect incident sunlight. 

\begin{figure}
\begin{center}
    \includegraphics[width=0.80\linewidth]{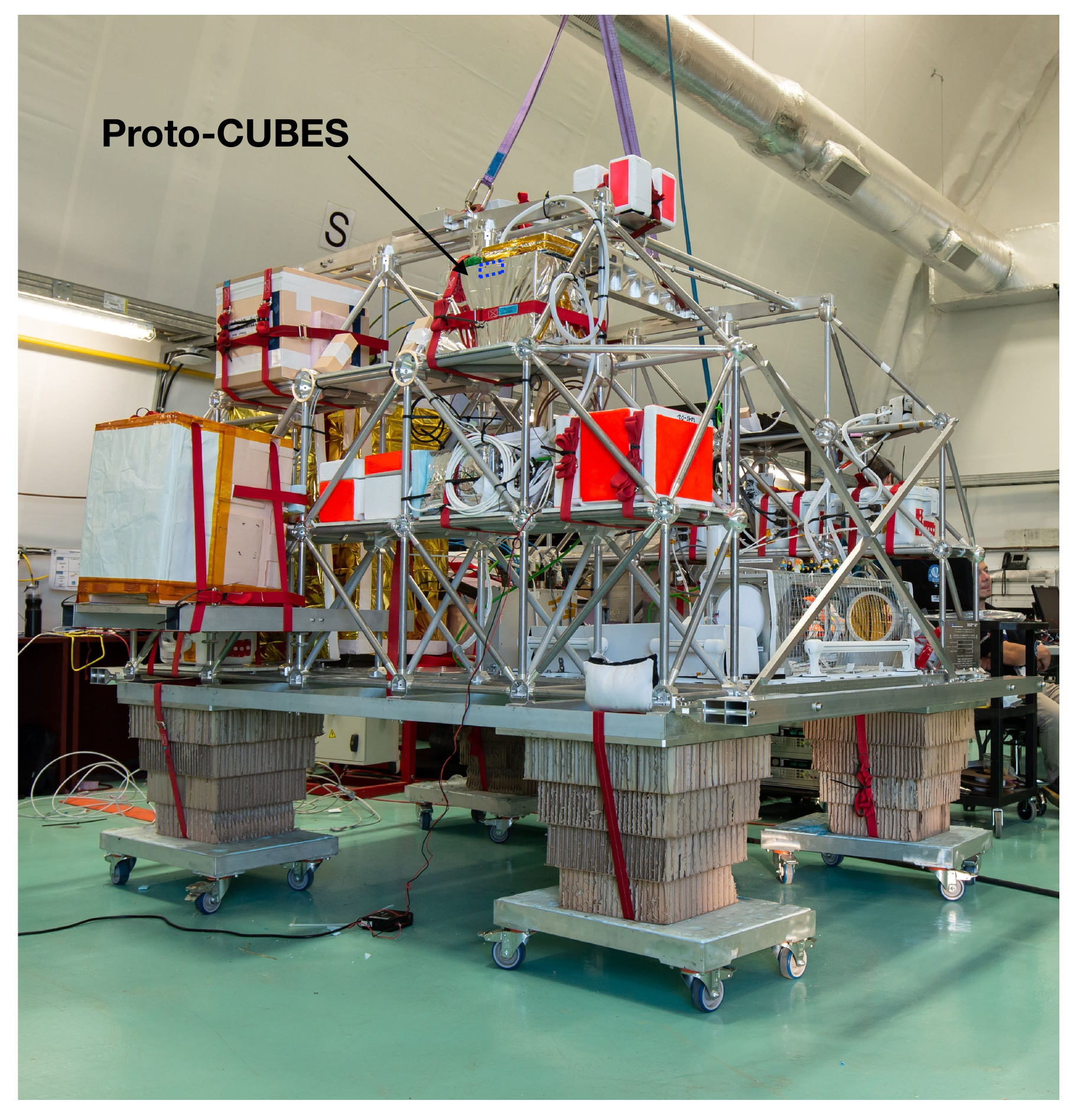}
\end{center}
\caption{Proto-CUBES installed on the CARMENCITA gondola prior to launch. The gondola structure stands $\sim$2.5~m tall, and weighs $\sim$420~kg. The hoisting straps connected to the top of the gondola illustrate how the gondola is connected to the balloon rigging. Four sacrificial cardboard crash pads are mounted under the gondola to lessen impact forces when the gondola lands by parachute after the flight. A number of payloads are distributed around the gondola (white boxes). Proto-CUBES is located at the top of the gondola (silver-coloured box).}
\label{fig:gondola} 
\end{figure}

The Proto-CUBES enclosure was installed on the French Space Agency (CNES) CARMENCITA gondola as part of the 
CABUX mission, which included a number of other payloads. 
Proto-CUBES was accommodated as part of the EU Research Infrastructure programme, HEMERA.
The location of the Proto-CUBES enclosure on the gondola is shown in Figure~\ref{fig:gondola}. The scintillator detectors have an unoccluded field-of-view over approximately the entire hemisphere directed away from the gondola perpendicular to zenith. X-rays entering the opposite hemisphere will be attenuated by gondola and payload materials.

\subsection{Balloon flight results}
The internal battery system was activated $\sim$30 minutes prior to launch.
Prior to this, a Am-241 source had been used to confirm that there was no significant change in the energy calibration for each MPPC channel between post-assembly testing in the laboratory in Stockholm and pre-flight tests at the launch site. 
The gondola was launched from Timmins Airport, Ontario, Canada (latitude: 48.568$^\circ$, longitude: -81.372$^\circ$), at 04:09 UT (23:09 local time) on 27th August 2019, suspended under a 150$\times$10$^3$m$^3$ zero pressure balloon. The gondola hung freely under the balloon with no azimuthal stabilisation. A planned flight termination occurred after $\sim$12~hours. The gondola subsequently landed safely by parachute in a forest (latitude: 48.471$^\circ$, longitude: -81.676$^\circ$).  

Data acquisition proceeded reliably through-out the flight. Pulse height histograms were transmitted to ground for each MPPC channel once per minute. The live-time registered for each histogram file was $>$95\% through-out the flight. The MPPC bias voltage was set to 54.5~V, and a temperature compensation of 50.9~mV/$^\circ$C was used. 

\begin{figure}
\begin{center}
    \includegraphics[width=0.90\linewidth]{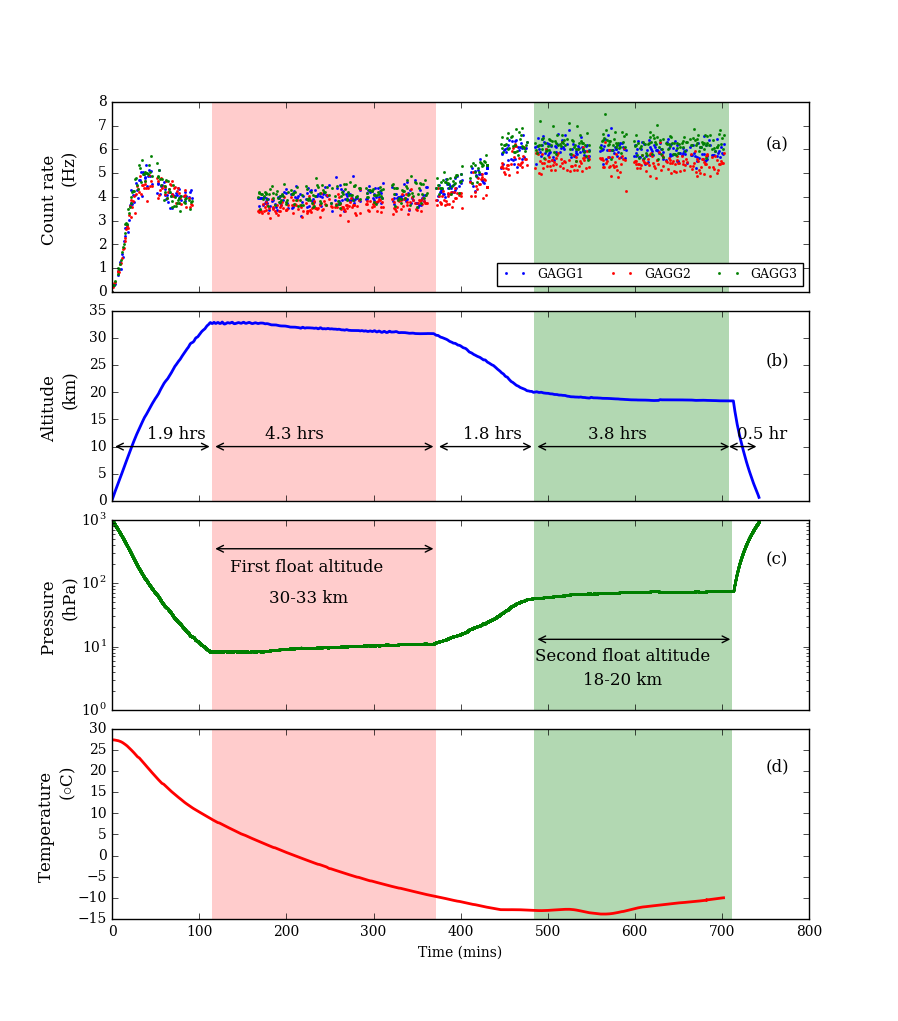}
\end{center}
\caption{The temporal variation of (a) count rate, (b) gondola altitude, (c) atmospheric pressure in the vicinity of the gondola, and (d) temperature inside the Proto-CUBES enclosure. Note that (a) and (d) show no data points at the end of the flight since the DAQ was stopped during the descent phase. Missing data points in the count rate plot arise when the detector was not configured for data-taking while other tests were conducted.}
\label{fig:timedep} 
\end{figure}

The time-dependence of key flight variables is shown in Figure~\ref{fig:timedep}. The flight comprised a $\sim$1.9~hour long ascent phase, two $\sim$4~hour long periods at approximately constant "float" altitude, $\sim$30~km (10~g/cm$^2$ vertical atmospheric overburden), and $\sim$20~km 
(80~g/cm$^2$), a $\sim$1.9~hour long transition between these altitudes, and a $\sim$0.5~hour long descent phase (mostly by parachute). Data collected during the float phases is the focus of this section. Gondola housekeeping data such as position, altitude, and atmospheric pressure were recorded by CNES during the flight. Temperature data was provided by a sensor inside the MPPC power supply module. The measured temperature covered the majority of the range expected in-orbit (see Section~\ref{sec:thermal}). A temperature below 0$^\circ$C (-10$^\circ$C) was recorded for $\sim$70\% (40~\%) of the flight. 

The temporal development of the counting rate obtained from triggered events in all three MPPCs is shown in Figure~\ref{fig:timedep}. 
During the ascent phase, the Pfotzer-Regener maximum is reached at an altitude of $\sim$20~km. The count rate measured during the ascent phase is less than that measured at a comparable altitude during the second float phase. This is because the altitude increases rapidly during ascent, resulting in a steady decrease in the instantaneous count rate after passing through the Pfotzer-Regener maximum, and consequently an underestimation of the counting rate during the 1 minute long data collection runs.
At an altitude of $\sim$33~km, the balloon altitude, and count rate, stabilised.
As expected, the counting rate correlates with atmospheric pressure after passing the Pfotzer-Regener maximum. 

\begin{figure}
\begin{center}
    \includegraphics[width=0.80\linewidth]{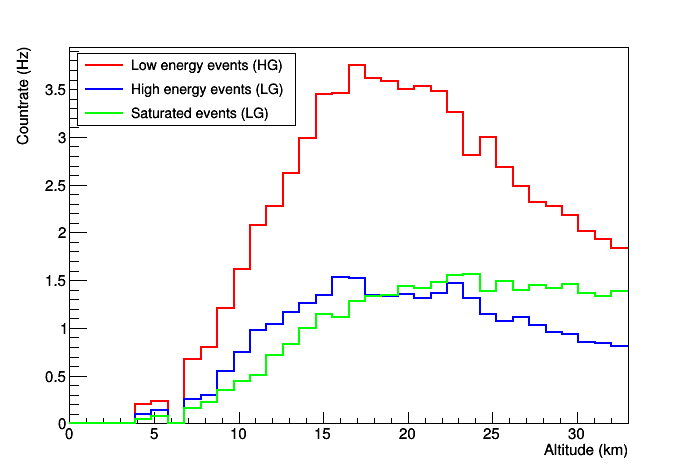}
\end{center}
\caption{The altitude dependence of count-rates for low-energy (high gain, HG), high-energy (low gain, LG), and saturated LG channels.}
\label{fig:altitudedep} 
\end{figure}

The dependence of the counting rate on atmospheric pressure for MPPC2 channel is shown in Figure~\ref{fig:altitudedep}, for low-energy (HG chain), high-energy (LG chain), and LG events which saturate the ADC connected to the output of the Citiroc. The saturated events are due to high energy photon interactions, as well as a significant contribution from   
minimum-ionising charged particles (MIPs), which deposit $\sim$5~MeV when traversing the full GAGG scintillator thickness of 5~mm.
The count-rate is dominated by low energy events, as expected due to the inverse power law nature of the incident X-ray radiation, and the energy dependence of the detection efficiency (discussed below). 

At higher altitudes, the count-rate for both LG and HG events decreases. The low-energy count-rate continually decreases. The high-energy rate starts to flatten at an altitude around 30 km, which is thought to be due to the high-energy component of the CXB scattering into the Proto-CUBES energy range. At this altitude (overburden 10 g/cm$^2$), CXB photons with an energy exceeding 50 keV have a transmission probability of $\sim$20\%~\cite{Dean_atmos}.
The count-rate for saturated LG events reaches a maximum value higher in the atmosphere (25~km, compared to 20~km for X-rays), and then stays approximately constant. Similar behaviour was reported~\cite{Lawrence} for measurements using other types of scintillator materials, Cs$_2$LiYCl$_6$ (CLYC) and CeBr$_3$, during a 22~day long stratospheric balloon flight around the Antarctic continent, and were subsequently replicated using a Monte Carlo simulation of cosmic-ray air showers.

\begin{figure}
\begin{center}
    \includegraphics[width=0.80\linewidth]{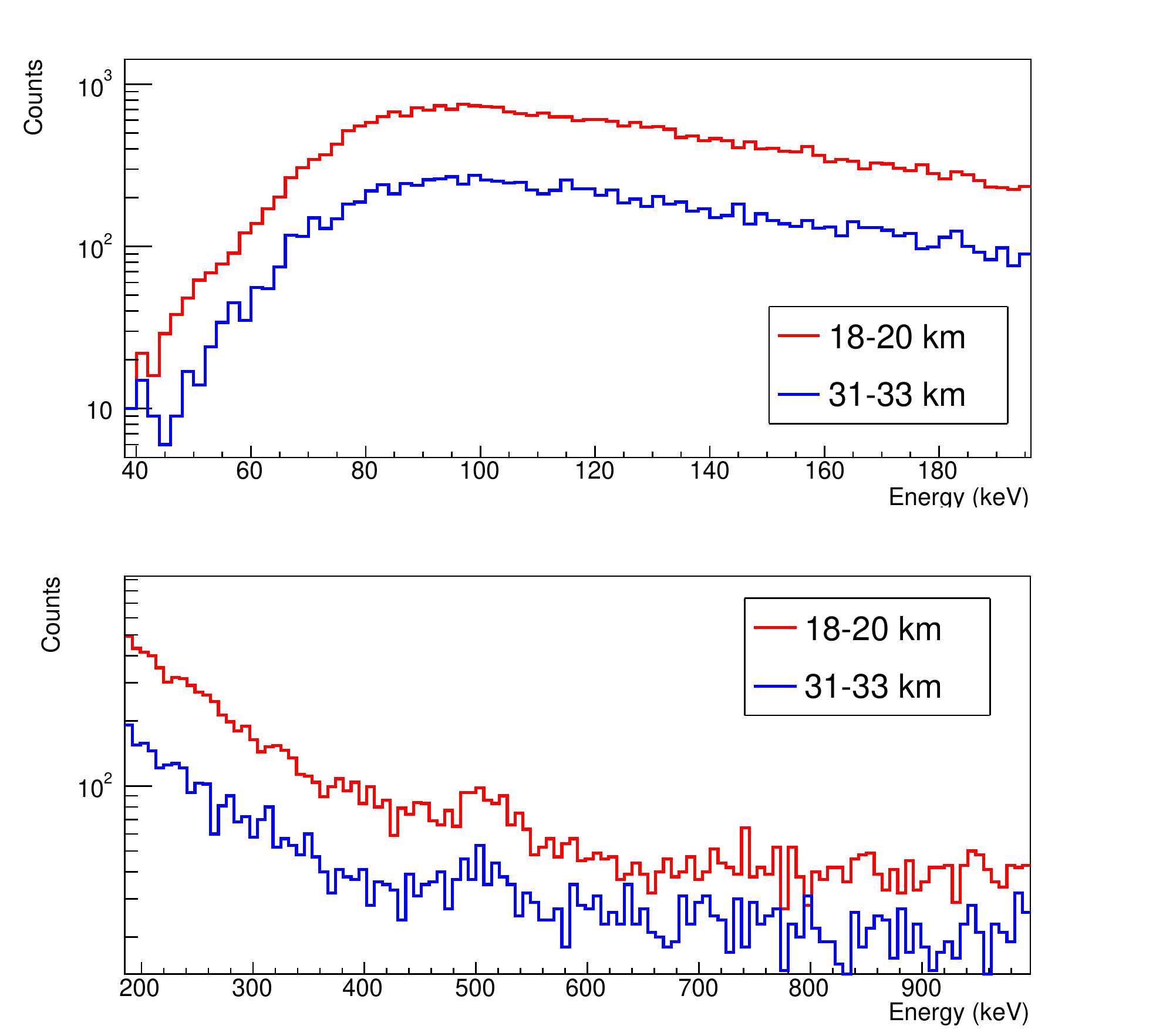}
\end{center}
\caption{The energy counts spectra for the HG and LG chains, evaluated at the two float altitudes.}
\label{fig:countsspectra} 
\end{figure}

The energy counts spectra recorded at the two float altitudes from the HG and LG chains for MPPC2 channel is shown in Figure~\ref{fig:countsspectra}. Aside from the normalisation, the spectral properties show no significant dependence on altitude. 
In order to convert the counts spectrum into an energy spectrum, the instrument response must be accounted for. This is beyond the scope of the work presented here. One component of the response is the GAGG scintillator X-ray detection efficiency. In~\cite{GAGGeff}, the $>$20~keV detection efficiency for a GAGG scintillator of similar dimensions to that used here is shown to decrease rapidly (photoelectric absorption dominates) until an incident energy of $\sim$500~keV, and then remain approximately constant (Compton scattering dominates) at $\sim$20\%. This tendency is seen in the LG channel. 
Since Proto-CUBES has a wide field-of-view, a detailed mass model of instrument and gondola components is also required, e.g. to account for attenuation and local particle production, as well as determining contributions from high-energy photons which Compton scatter into the Proto-CUBES energy range. 

\begin{figure}
    \begin{center}
        \includegraphics[width=0.80\linewidth]{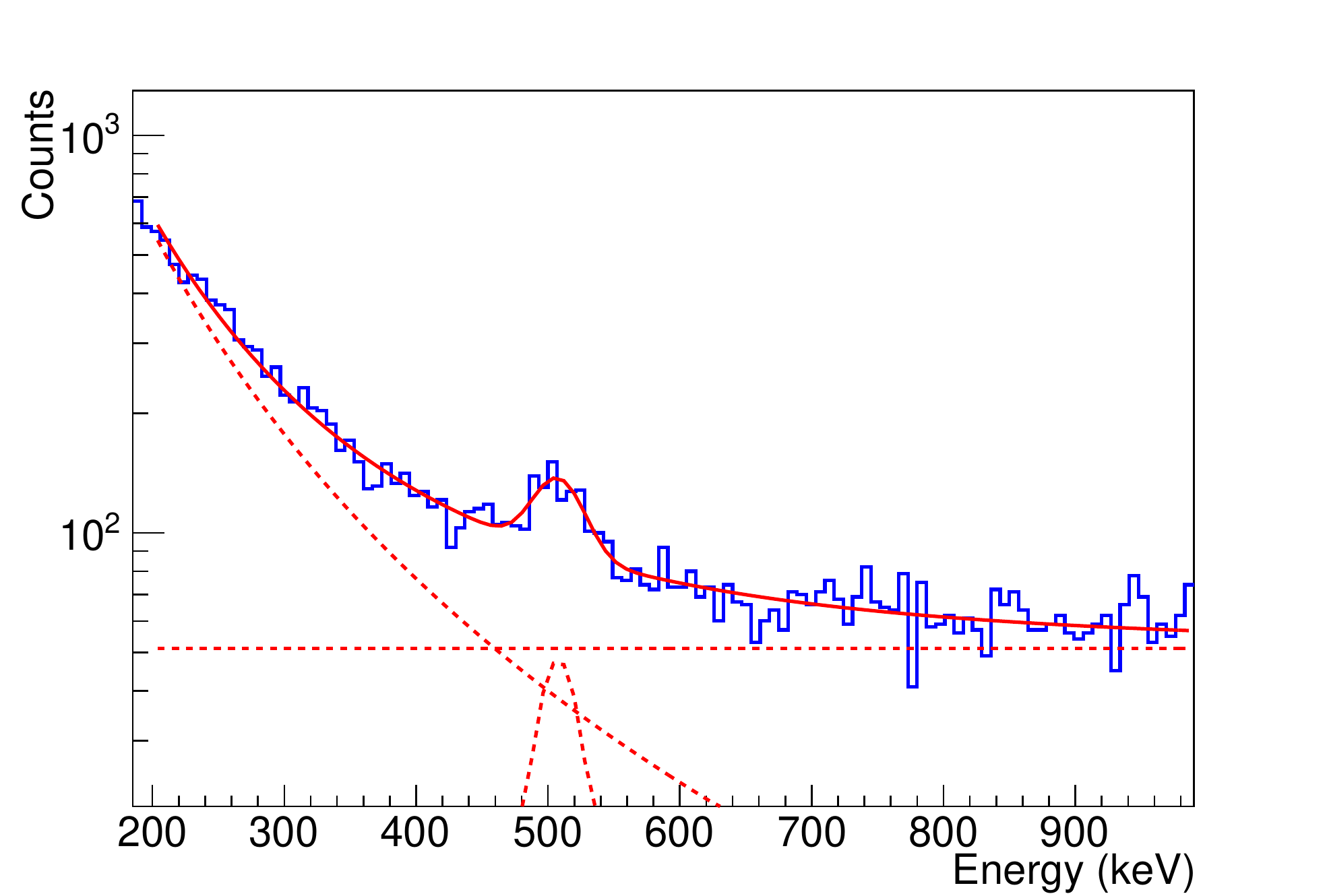}
    \end{center}
    \caption{The summed energy counts spectrum for two float altitudes (from the LG channel). The summed spectrum is fit with $a\left(\frac{E}{300}\right)^{b} + c + d\exp(-0.5((E - e)/f)^{2}$), where E is the energy in keV. The normalisation, $a$ is specified at 300 keV. The fitted parameters are listed in Table~\ref{tab:fit_par_fig18}.}
    \label{fig:addedspectra} 
\end{figure}

In Figure~\ref{fig:countsspectra}, the HG channel shows a low-energy cut-off around 40~keV, which is dictated by the Citiroc trigger threshold. The suppression of counts in the energy range 40--90~keV is a consequence of how the Citiroc trigger system is implemented. Due to the nature of the tests with radioactive sources (Section~\ref{sec:energy-range}), this effect was not evident prior to the balloon flight, but it will need to be characterised, and corrected for in subsequent flight data.  
In the LG channel, an excess of events is evident at 511~keV. This is more clearly seen in Figure~\ref{fig:addedspectra}, where the counts spectra acquired at both float altitudes are summed. The summed spectrum is well described by a power law, a Gaussian feature at 511~keV and a constant.The parameters from the fit are listed in Table~\ref{tab:fit_par_fig18}. The 511 keV energy resolution obtained in flight, (8.5$\pm$0.2)\%, is smaller than that obtained during laboratory tests conducted at room temperature, (11.7$\pm$0.2)\% (Figure~\ref{fig:resoln}). This discrepancy may be due to the lower temperature experience by the GAGG scintillator in flight, 0 - -15$^{\circ}$C. This trend matches that reported elsewhere~\cite{yoneyama}, although the energy resolution is reported to be temperature independent in other work~\cite{gaggtemp}, so further study is warranted.

The Gaussian feature at 511 keV is ascribed to a positron annihilation line smeared by the GAGG scintillator energy resolution.
Such a line may arise {\it (i)} if a primary or secondary cosmic-ray position interacts with material in the vicinity of the scintillators; {\it (ii)} due to the decay of spallation products (e.g. Na-22) created when cosmic-ray protons interact in the aluminium materials surrounding Proto-CUBES; {\it (iii)} from pair production by high energy photons which interact in dense materials surrounding Proto-CUBES. An intrinsic 511~keV feature may also arise due to cosmic-ray proton interacting in the GAGG scintillator, which leads to the $\beta^+$ decay of Gd (51\% mass fraction in GAGG). The presence of Gd also allows for the capture of thermal neutrons, with associated production of X-ray lines~\cite{Taggart, GAGGeff}. While several lines are expected to fall in the Proto-CUBES energy range, these were not observed during the relatively short balloon flight. Independent of origin, the 511~keV feature provides a convenient way to monitor the energy scale calibration during flight. 

    \begin{table}[H]
        \centering
        \caption{\label{tab:fit_par_fig18} Fit parameters of $a\left(\frac{E}{300}\right)^{b} + c + d\exp(-0.5((E - e)/f)^{2}$ in Figure~\ref{fig:addedspectra}}
        \smallskip
        \begin{tabular}{|c|c|c|}
        \hline
        Parameter&Value\\
        \hline
        $a$ (counts)&177.04$\pm$4.03\\
        \hline
        $b$&-2.91$\pm$0.08\\
        \hline
        $c$ (counts)&51.21$\pm$1.67\\
        \hline
        $d$ (counts)&47.85$\pm$6.72\\
        \hline
        $e$ (keV)&507.77$\pm$2.83\\
        \hline
        $f$ (keV)&18.59$\pm$2.77\\
        \hline
        \end{tabular}
    \end{table}

\section{Outlook}
\label{sec:outlook}

\begin{figure}[H]
  \centerline{\includegraphics[width=0.8\textwidth]{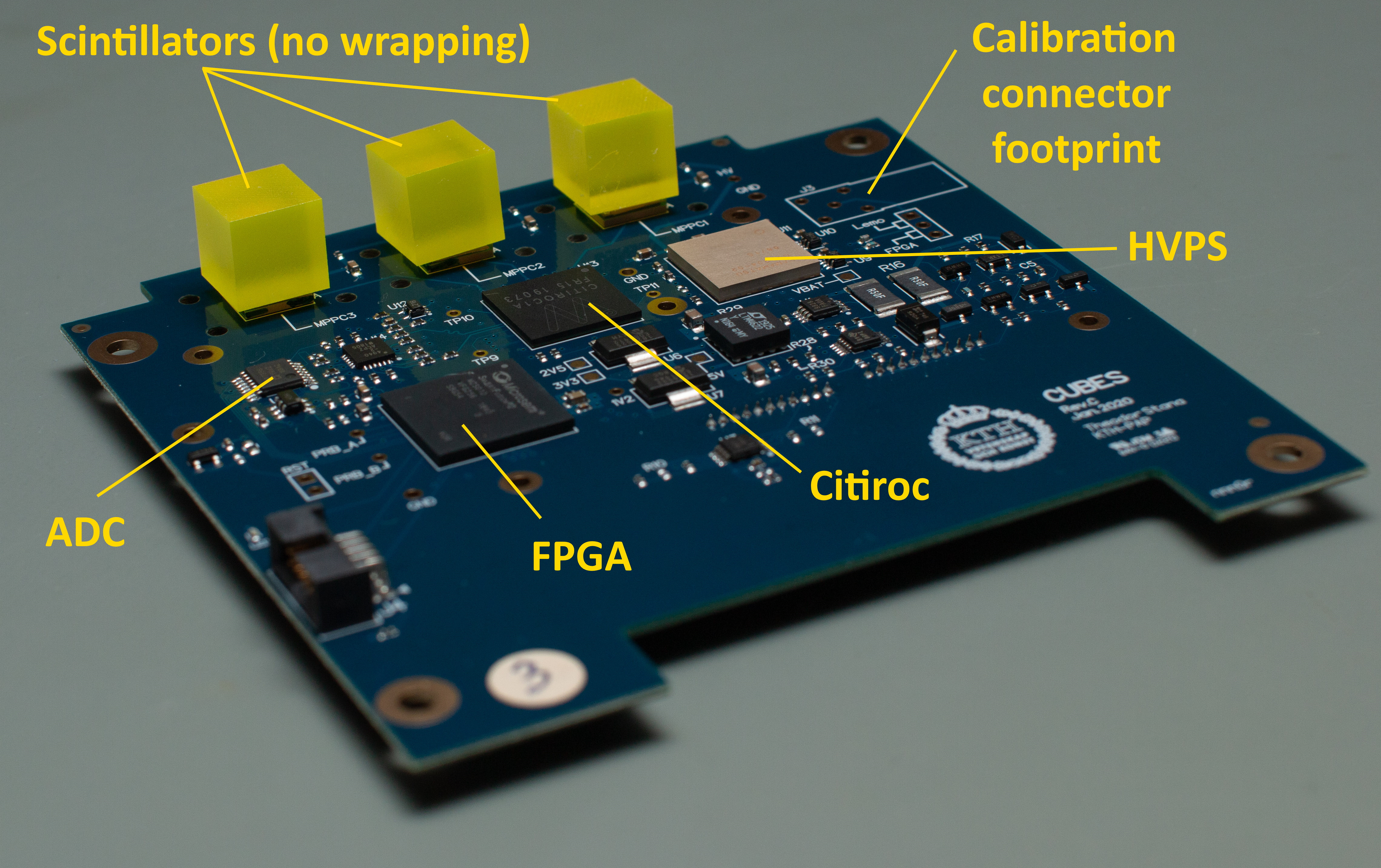}}
  \caption{The CUBES flight PCB (95~$\times$~95~mm). The GAGG scintillator housing is not shown. The upper and lower surfaces of the PCB are covered in a $\sim$1~mm thick aluminium sheet to provide a heat path for components, and to shield components from low energy cosmic ray electrons.}
  \label{fig:flightpcb}
\end{figure}

When work on CUBES started, it was foreseen that a single detector board, of the size shown in Figure~\ref{fig:cubes-pcb-revB}, would be installed on MIST. 
As the development work for CUBES progressed, the opportunity to install two PC/104-format CUBES detectors boards arose.
Following the Proto-CUBES balloon flight, a new version of the CUBES PCB was therefore designed. 
The new CUBES PCB, shown in Figure~\ref{fig:flightpcb}, is approximately twice the size of the PCB used for Proto-CUBES, but the core components are unchanged.
Characterisation work on the new CUBES PCBs is on-going, using the same
test setup as shown in Figure~\ref{fig:cubes-block-diag-dev-setup}.

The Citiroc ASIC provides a dedicated calibration input, but this was not implemented for Proto-CUBES. 
During the nominal one year mission, GAGG scintillator and/or MPPC characteristics may
change due to exposure to cosmic radiation. To allow the ASIC response to be studied independently from these components,  
the Citiroc calibration input has been connected to an
output on the FPGA via a 100~pF capacitor. The FPGA can drive a 2.5~V pulse
through the capacitor to generate a 250~pC charge pulse, there-by simulating a MPPC signal. 
During on-the-ground
characterisation of the instrument, a pulse generator can be connected to the
calibration input to characterise the linearity of the ASIC without using the FPGA. 

The $\sim$25~kB histograms stored on Proto-CUBES will be too large for the
limited data downlink capabilities provided within MIST. As a result, the DAQ system in the flight
version of CUBES will compress the data prior to sending it to the OBC for
downlink. Several compression types are being developed, from fixed-width
rebinning to logarithmic scale rebinning. Although the data will be re-binned
by CUBES prior to sending to the OBC, data during a DAQ run will still be stored
at full size, i.e., 2048 bins per histogram. It is foreseen that the full-size
bins will be downlinked during the early stages of operation, while  commissioning CUBES.

\acknowledgments
Part of this work derives from the Phase A studies for the SPHiNX mission, which was funded by the Swedish National Space Agency (grant number 232/16).
MP acknowledges support from the Swedish Research Council (grant number 2016-049929). 
The Proto-CUBES balloon flight was provided through the EU Horizon 2020 Research Infrastructure programme, HEMERA. 
As well as providing a flight opportunity, funding was awarded through the HEMERA programme to allow KTH personnel to participate in the balloon campaign. 
Personnel from Centre National d'Etudes Spatiales (CNES), and, in particular, Fr\'{e}d\'{e}ric Blon, Nicolas Bray, St\'{e}phane Louvel, 
and Andr\'{e} Vagas, are gratefully acknowledged for their assistance with technical and logistical practicalities surrounding the balloon flight.
Thermal and vacuum test facilities were provided by the KTH Space Centre. The MIST team, under the leadership of Sven Grahn, is thanked for providing the interface details required to develop CUBES.
Marcus Persson is thanked for his contributions to the development of the Proto-CUBES data acquisition system, and for his participation in the balloon flight campaign. The authors are very grateful to Hiromitsu Takahashi (Hiroshima University) for providing the GAGG scintillators used in this work. 

\bibliography{referencesJINST}
\bibliographystyle{JHEP}

%
%
 %



\end{document}